\documentclass[aps,twocolumn,groupedaddress,showpacs,superscriptaddress]{revtex4-1}

\usepackage{graphicx}
\usepackage{float,amsmath,verbatim,amssymb,stmaryrd}
\usepackage{epsfig}
\usepackage{pifont}

\begin{document}

%\preprint{}

\title{Seeded QED cascades in counter propagating laser pulses}

\author{T. Grismayer }
\email{thomas.grismayer@ist.utl.pt}
\address{GoLP/Instituto de Plasmas e Fus\~ao Nuclear, Universidade de Lisboa, Lisbon, Portugal}
\author{M. Vranic}
\address{GoLP/Instituto de Plasmas e Fus\~ao Nuclear, Universidade de Lisboa, Lisbon, Portugal}
\author{J. L. Martins}
\address{GoLP/Instituto de Plasmas e Fus\~ao Nuclear, Universidade de Lisboa, Lisbon, Portugal}
\author{R. A. Fonseca}
\address{GoLP/Instituto de Plasmas e Fus\~ao Nuclear, Universidade de Lisboa, Lisbon, Portugal}
\address{DCTI/ISCTE Instituto Universit\'{a}rio de Lisboa, 1649-026 Lisboa, Portugal}
\author{L. O. Silva}
\email{luis.silva@ist.utl.pt}
\address{GoLP/Instituto de Plasmas e Fus\~ao Nuclear, Universidade de Lisboa, Lisbon, Portugal}

\date{\today}

\begin{abstract}

The growth rates of seeded QED cascades in counter propagating lasers are calculated with first principles 2D/3D QED-PIC simulations. The dependence of the growth rate on laser polarization and intensity are compared with analytical models that support the findings of the simulations. The models provide an insight regarding the qualitative trend of the cascade growth when the intensity of the laser field is varied. A discussion about the cascade's threshold is included, based on the analytical and numerical results. These results show that relativistic pair plasmas and efficient conversion from laser photons to gamma rays can be observed with the typical intensities planned to operate on future ultra-intense laser facilities such as ELI or VULCAN. 

 \end{abstract}


%\keywords{}

\maketitle
%Intro_

\section{Introduction}

The process of electron-positron pair creation from photon decay has been known since the early 30's but only the striking E-144 SLAC experiment \cite{E144slac1,E144slac2} first demonstrated the possibility of producing matter directly via light-by-light scattering. The limits of the laser technology ($I\sim 10^{19}$ W/cm$^2$) at the time constrained the experiments to use the ultra relativistic SLAC electron beam in order to reach the quantum electrodynamic (QED) regime necessary for the observation of pair production. The recent spectacular rise in laser intensities, accompanied by the ongoing construction of new laser facilities such as ELI \cite{ELI} or the Vulcan 20 PW Project \cite{Vulcan} will place intensities above $10^{23}$ W/cm$^2$ within reach thus allowing for the exploration of new physics regimes \cite{RevModPhysdipiazza}. Different laser configurations that have been envisaged to lower the intensity threshold \cite{ Bulanov2006} in order to observe Schwinger-like pair creation but prolific vacuum pair production requires higher intensities than these aforementioned. Therefore, one ought to consider pair creation through the decay of high energy photons in intense fields. This process usually leads to QED cascades, as the pairs created reemit hard photons that decay anew in pairs, eventually resulting in an electron-positron-photon plasma. QED cascades, also referenced as electronic or electromagnetic showers \cite {Landau_showers, Akhiezer, Anguelov_Vankov, Erber} when the external field is purely magnetic, have been theoretically studied in different electromagnetic configurations \cite{Bulanov_pairsvaccuum, Bell_Kirk_2lasers, Bulanov2013, Gelfer}. Notably, Bell \& Kirk \cite{model1bell} suggested a judicious configuration comprising two circularly polarized counter propagating lasers with some electrons in the interaction region to seed the cascade. They predicted prolific pair production for intensities approaching $10^{24}$ W/cm$^2$ for a $\mu$m wavelength laser. Recently several groups \cite{Nerush_laserlimit, Bashmakov, Jirka} have pioneered the investigation of such cascades in the counter propagating laser setup with particle-in-cell simulations where a QED phenomena such as photon emission and pair creation have been added. In this paper we intend to qualitatively determine the conditions under which driven cascades operate. We resort to 2D/3D QED-PIC simulations to calculate their associated growth rates for different laser polarizations and for a wide range of intensities. The numerical results are then compared to an analytical model in two asymptotic limits. This model opens the way for determining the optimal conditions to generate dense electron-positron plasmas in the laboratory. 

\section{Simulations}

Our exploration relies on a QED module, part of our particle-in-cell (PIC) code OSIRIS 3.0 \cite{OSIRIS}, which includes real photon emission from an electron or a positron, and decay of photons into pairs, the Breit-Wheeler process.  The differential probability rates describing these processes can be found in \cite{pair_rate1, pair_rate2, pair_rate3, Ritus_thesis, Erber}. The implementation of such a module has already been described in detail elsewhere \cite{Gremillet, Ridgers_quantumRRnew, Elkina_rot,Ridgers_solid, Nerush_laserlimit, Bell_Kirk_MC}. Many QED-PIC simulations have been performed in order to benchmark our module with previous results \cite{Gremillet, Elkina_rot, Nerush_laserlimit, Bashmakov, Tang}. The exponential growth of the number of PIC particles, a critical numerical issue, is sorted out with the use of a novel particle merging algorithm \cite{Vranic_merging} that resamples the 6D phase space with different weighted macro-particles, allowing parameter scans in 2D/3D. The algorithm preserves the total energy, momentum and charge as well as the particle phase space distribution  whereas previous attempts to merge particles were only focused on the conservation of some of the physical quantities \cite{Timokhin, Lapenta_new, Lapenta_old}.

\begin{figure*}%\centering
\includegraphics[width=1\textwidth]{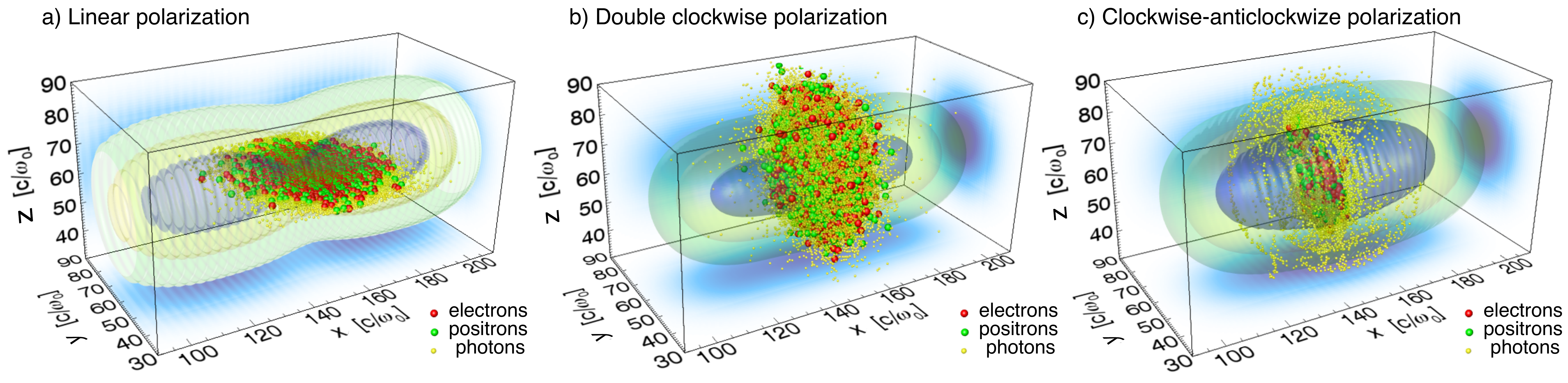}
\caption{3D PIC simulation snapshot of QED cascades for a) Setup 1 with $a_0=1000$ at $t=90~\omega_0^{-1}$, b) Setup 2 with $a_0=1300$ at $t=80~\omega_0^{-1}$, c) Setup 3 with $a_0=2000$ at $t=46~\omega_0^{-1}$. The laser pulses are shown through iso-contours of the electromagnetic energy. The particles displayed represent only a small fraction of the simulation particles.} 
\label{fig:cascade3d}
\end{figure*}

To motivate our discussion we first present simulations where we have explored three configurations of colliding laser pulses whose polarization can either be linear or circular. The three-dimensional development of the cascade is shown in Fig.\ref{fig:cascade3d} for different configuration. By examining the geometry of the standing waves, we can develop an intuitive picture on how the particles are accelerated, and hence predict which configuration will be optimal. For a given $a_0$, the optimal configuration consists in favoring the maximal pair growth and hence determining which field configuration offers on average the highest values of $\chi = (e\hbar/m^3c^4)\sqrt{(\gamma\vec{E}+\vec{u}\times \vec{B})^2-(\vec{u}\cdot\vec{E})^2}$ with $\vec{u}=\vec{p}/mc$. It should be emphasized that radiation reaction in intense fields modifies the orbits of particles \cite{Esirkepov} and can lead to anomalous radiative trapping \cite{Gonoskov2014} which we omit in the following analysis but which is self-consistently captured in our simulations.
 
Setup 1 (lp-lp) consists of two linearly polarized lasers where the phase and polarization are defined by 
\begin{equation}\label{first_wave}
\vec{a}_\pm=(0,a_0 \cos(\omega_0t \pm k_0x),0), 
\end{equation}
where ``$-$'' and ``$+$'' respectively denote a wave propagating in the positive and in the negative $x$ direction. $a_0=eE_0/m\omega_0c$ is the Lorentz-invariant parameter, related to the intensity $I$ by $a_0=0.85(I\lambda_0^2/10^{18}\textrm{Wcm}^{-2})^{1/2}$ and $E_0$ the peak electric field strength (which will be expressed is units of $m\omega_0 c /e$ in the following, such that $\tilde{E}_0=eE_0/m\omega_0c = a_0$). This results in a standing wave where $E_y=2a_0 \cos (k_0x) \sin(\omega_0t),  B_z=-2a_0 \sin (k_0x) \cos (\omega_0t)$; the electric and magnetic fields of the standing wave have a fixed direction, and $\vec{E}\perp \vec{B}$ and there is a $\pi/2$ phase offset between $\vec{E}$ and $\vec{B}$ both in space and time. This hints that the dynamics of the particles in the standing wave might be dominantly affected by the electric or the magnetic field depending on the phase within the temporal cycle \cite{Bashmakov, Esirkepov}. The electric field accelerates electrons in the $y$ direction, while the magnetic field $B_z$ can rotate the momentum vector and produce also $p_x$ and the orbits are confined in the $x-y$ plane, see Fig.\ref{fig:cascade3d}a). The existence of the $p_x$ component ensures that there is a perpendicular momentum component to both $\vec{E}$ and $\vec{B}$. Rotating the momentum vector towards the higher $p_x$ gradually increases $\chi_e$ until a photon is radiated. This photon then propagates and can decay far form the emission point. For a particle born at rest, $\chi_e$ oscillates approximatively twice per laser period with a maximum on the order of $2a_0^2/a_S$ where $a_S=mc^2/\hbar\omega_0$ is the normalized Schwinger field \cite{Schwinger}. The cascade develops mostly around the bunching locations (two per wavelength that corresponds to the moment of rotation or high $\chi$) and is characterized by a growth rate that possesses an oscillating component at $2\omega_0$.  

Setup 2  (cw-cw) is composed of two clockwise circularly polarized lasers defined by 
\begin{equation}
\vec{a}_\pm=(0,a_0\cos(\omega_0t\pm k_0x),\pm a_0\sin (\omega_0t\pm k_0x)),
\end{equation}
where $a_0=0.6(I\lambda_0^2/10^{18}\textrm{Wcm}^{-2})^{1/2}$. In addition to the $E_y$ and $B_z$ components that are the same as for the lp-lp case, we also have $E_z=2a_0 \sin (k_0x) \sin(\omega_0t), B_y=-2a_0 \cos (k_0x) \cos (\omega_0t)$. For any $x$, both $\vec{E}$ and $\vec{B}$ are parallel to the vector $\vec{e}(x)=(0,\cos x, \sin x)$. The direction of the fields depends on the position, but the amplitude of both $\vec{E}$ and $\vec{B}$ is only a function of time,  which results in a helical field structure growing or shrinking uniformly in space ($\vec{E}$ and $\vec{B}$ are dephazed by $\pi/2$ in space). Contrary to the lp-lp setup, this configuration does not produce $p_x$ for particles born at rest since at each position both $\vec{E}$ and $\vec{B}$ are parallel to the momentum at all times, and significant $\chi_e$ cannot be achieved. Reaching high values of $\chi_e$ is, however, possible for particles that are not at rest initially. If an external perturbation provides a transverse momentum $p_x$ (e.g. the initial ponderomotive force due to the laser pulse envelope) the particle can move along the $x$ axis and leave the region where the fields remain parallel to the momentum kick acquired at the initial position. In this way the value of $\chi_e$ is increased, and so is the probability of radiating hard photons. The decay of hard photons produces pairs that will either possess an initial transverse or longitudinal momentum component and the cascade will naturally develop. A crude analysis shows that the maximal $\chi_e$ attainable is on the order of $2a_0\gamma_0/a_S$ ($\gamma_0$ being the initial energy of the particle when created). All $x$ positions have equivalent probabilities to initiate a cascade because only the azimuthal angle of the field changes along the  $x$ axis. Therefore, the cascade shall develop over the entire wavelength. 

Setup 3  (cw-cp) is formed by a clockwise and a counter-clockwise polarized laser: 
\begin{equation}
\vec{a}_\pm=(0,a_0\cos(\omega_0t\pm k_0x),- a_0\sin (\omega_0t\pm k_0x)),
\end{equation} 
where $a_0=0.6(I\lambda_0^2/10^{18}\textrm{Wcm}^{-2})^{1/2}$. The components $E_y$ and $B_z$ are anew the same but $E_z=2a_0 \cos(k_0x) \cos(\omega_0t), B_y=-2a_0 \sin(k_0x) \sin(\omega_0t)$. The magnitude of the field vectors is constant in time ($|\vec{E}|=2 a_0 \cos x$ and  $|\vec{B}|=2 a_0 \sin x$) whereas the direction changes. In this case, $\vec{E} || \vec{B}$, and their direction $\vec{e}(t)=(0, \cos t, \sin t)$ does not depend on space, which results in a fixed planar beating pattern which rotates around the laser propagation axis. This setup consists in a rotating field structure and the dynamics of the particles has been already studied \cite{model1bell,Elkina_rot,Fedotov_cascade}. The advantage lies in the direction of the fields that is constantly changing, and the particles are not required to move in $x$ to enter a region where  $\vec{E}$ and $\vec{B}$ are perpendicular to their momentum. For similar $p_\perp$, the $\chi_e$ is on the same order regardless of the $x$ position, so we could expect the cascade to grow everywhere with the same probability. However, the particle acceleration is stronger in the regions of high electric field, so the highest electron momenta are obtained where the electric field is maximum. This then leads to higher $\chi_e$ and the cascade favorably develops in the region of strong electric field (precisely in the node where $B=0$ \cite{model1bell}) producing a plasma wheel as shown in Fig.\ref{fig:cascade3d}c). At this particular position, the parameter $\chi_e$ can reach a maximal value of $2a_0^2/a_S$ \cite{Elkina_rot}.

From the description of the three configurations it seems clear that the second setup can be considered as non optimal in view of the low values of $\chi_e$. Rigorously, setup 1 can produce the highest values of $\chi_e$ ($\chi_e > 2a_0^2/a_S$) but only for particles born in a specific phase of the standing wave. The majority of the particles are sloshing back and forth between the electric and magnetic zone which results in lower average $\chi_e$ in comparison with setup 3. The efficiency of the cascade setups can be more accurately assessed by computing its growth rate $\Gamma$. We measure the growth rate in simulations, and compare it to the analytical prediction when possible. As a matter of fact, full analytical treatment is not always possible due to the complexity of the stochastic orbits in the standing wave. We introduce here models that allow to retrieve asymptotic limit of the growth rate. 

\section{Cascade models}

\subsection{Ideal model}

For a collection of identical photons $n_{\gamma}$, whose probability rate to decay into a pair is given by $W_p$, the number of pairs created after a time $t$ is $n_{p}=n_{\gamma}(1-e^{-W_{p}t})$. If the photons originate from a source that emits constantly at a rate $W_{\gamma}$, we get: $n_{p}=\int_0^tdt'W_{\gamma}(1-e^{-W_{p}(t-t')})$. The rate of created pairs is then
\begin{equation}
\frac{dn_{p}}{dt}=\int_0^tdt'W_{\gamma}W_{p}e^{-W_{p}(t-t')}
\end{equation}
If the source of the emitted photons are the pairs, the number of photons created during a time $dt'$ being $2dt'W_{\gamma}$ has to me multiplied be the current number of pairs $n_p(t')$. The rate of pairs is now
\begin{equation}
\label{eq:cascade1}
\frac{dn_{p}}{dt}=2\int_0^tdt'n_{p}(t')W_{\gamma}W_{p}e^{-W_{p}(t-t')}
\end{equation}
This equation can be solved using the Laplace transform. Defining the Laplace variable as $s$, Eq.(\ref{eq:cascade1}) becomes
\begin{equation}
\hat{n}(s)=\frac{n(0)}{s-\frac{2W_{\gamma}W_p}{s+W_p}}. 
\end{equation}
 The behavior of $n_p$, defined as the inverse Laplace transform of $\hat{n}$ depends at late times, $t \gg W_{\gamma}^{-1},W_p^{-1}$, on the contribution of the singularities of $\hat{n}$. These singularities are the roots of the polynomial $s^2+W_{p}s-2W_{\gamma}W_p=0$ that admits a positive and a negative solution
\begin{equation}
s_{\pm}=\frac{W_p}{2}\left(-1\pm\sqrt{1+8\frac{W_{\gamma}}{W_p}}\right),
\end{equation}
where the positive solution $s_{+}$ is consistent with result derived by Bashmakov\cite{Bashmakov}. A pole at $s_{\pm}$ gives a contribution scaling as  $e^{s_{\pm}t}$, thus the function $n_{p}(t)$ grows exponentially with a growth rate $\Gamma = s_{+}$. It is interesting to look how the growth rate depends on the characteristic rate as the ratio $r=W_{\gamma}/W_p$ between them evolves. 
\begin{equation}
\Gamma \simeq 
\begin{cases} 
2W_{\gamma} &\mbox{if } r \ll 1 \\
W_p\sim W_{\gamma} & \mbox{if } r \sim 1 \\
\sqrt{2W_{\gamma} W_p} & \mbox{if } r\gg 1
 \end{cases} 
\end{equation}

\subsection{Rotating field model}

\begin{figure}
\includegraphics[width=0.5\textwidth]{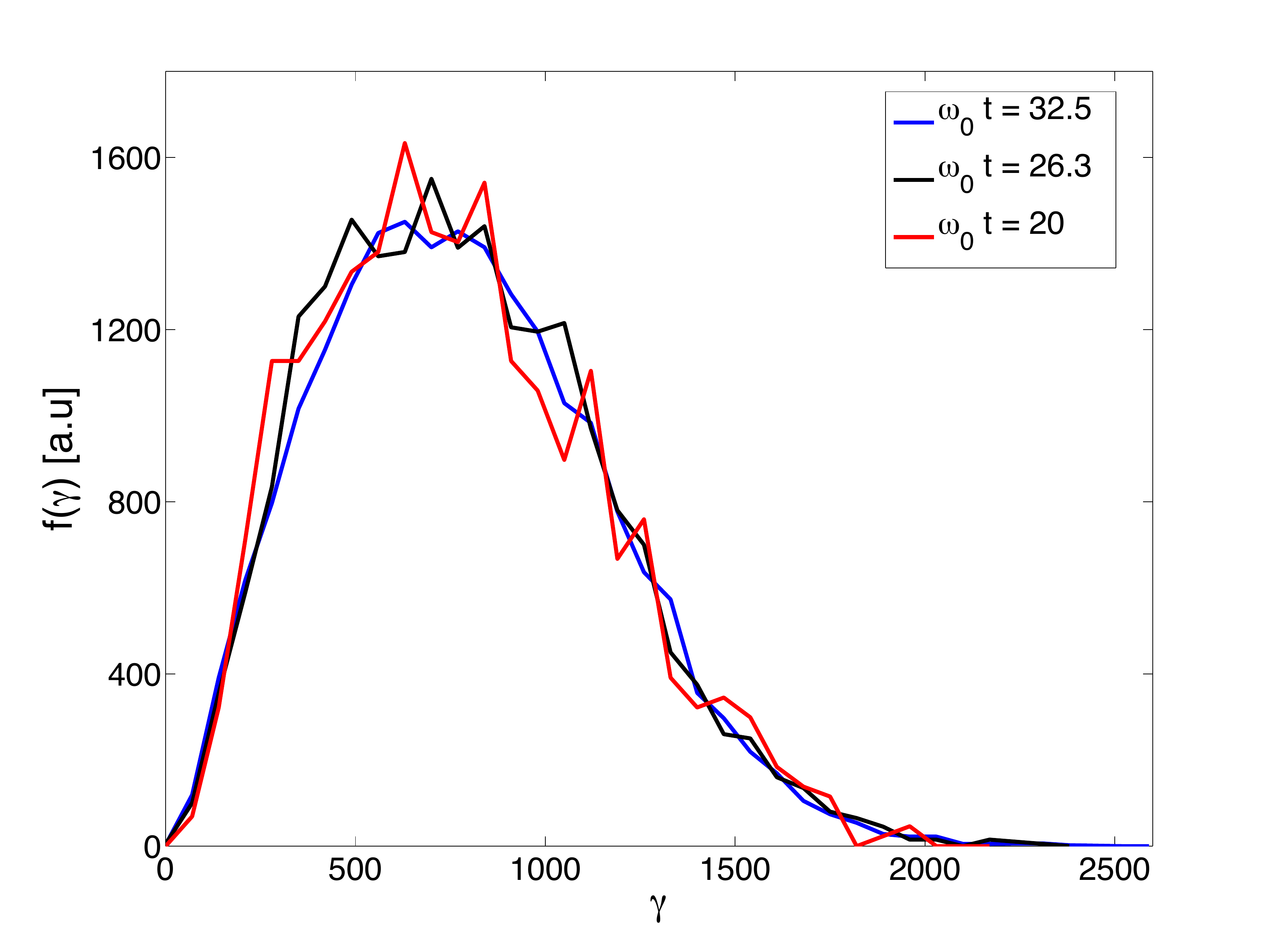}
\caption {Electron energy distribution function at different times.} 
\label{distribution}
\end{figure}

The case of a uniform rotating electric field constitutes a good approximation of the standing wave field produced in the setup 3  \cite{model1bell}. The advantage of this setup is that the cascade develops mostly in one spot $x=0$, which allows us to assume a time-depending field. In a full self-consistent kinetic model, we would have to follow all particles including quantum radiation reaction for their stochastic orbits, which appears to be overly cumbersome. Since the growth of the cascading process in a uniform rotating electric field is purely exponential \cite{Elkina_rot}, this means somehow that the electron (or positron) energy distribution remains almost constant during the development of the cascade (the number of particles increase but the shape of the distribution is not altered). This implies that in the tail of the distribution, the high-energy particles, that lose energy emitting photons, are constantly replaced by new created particles being reaccelerated. Figure(\ref{distribution}) shows the electron energy distribution at three different times (the details of the simulation parameters are thoroughly discussed in Sec.(\ref{section_para})). These three times are taken at consecutive rotation period of the electric field and one can notice that the shape of the distribution remains constant as the number of pairs grows in the system. We can then suppose that the pairs follow a fluid-like behavior which can be described through an average energy $\bar{\gamma}$ and an average quantum parameter $\bar{\chi_e}$. We generalize Eq.(\ref{eq:cascade1}) in the following form
\begin{equation}
\label{eq:intdiff}
\frac{dn_{p}}{dt}=2\int_0^tdt'\int d\chi_{\gamma}n_{p}(t')\frac{d^2P}{dt'd\chi_{\gamma}}W_{p}e^{-W_{p}(t-t')}
\end{equation}
The differential probability rate $d^2P/dt'd\chi_{\gamma}$ depends thus on $\bar{\gamma}$, $\bar{\chi}_e$ and $\chi_{\gamma}$. We further assume that the photon decay rate (or the pair emission probability rate) can be considered as constant in time which permits us to write $W_p=W_p(\chi_{\gamma},\epsilon_{\gamma})$ with $\epsilon_{\gamma}=\bar{\gamma}\chi_{\gamma}/\bar{\chi}_e$. Eq.(\ref{eq:intdiff}) is solved in the same way as before for Eq.(\ref{eq:cascade1}) and calculating $\Gamma$ amounts to solving the zeros of 
\begin{equation}
\label{eq:charac}
s-2\int_0^{\bar{\chi_e}} d\chi_{\gamma}\frac{\frac{d^2P}{dt'd\chi_{\gamma}}W_{p}}{s+W_p} = 0
\end{equation}

{\it Weak field limit:} when $\bar{\chi}_e\ll 1$, the pair creation probability can be approximated by \cite{Erber,Ritus_thesis} $W_p\simeq(3\pi/50)(\alpha/\tau_c)e^{-8/3\chi_{\gamma}}\chi_{\gamma}/\epsilon_{\gamma}$ and $d^2P/dtd\chi_{\gamma}\simeq \sqrt{2/3\pi}(\alpha/\tau_c)e^{-\delta}/(\delta^{1/2}\bar{\chi}_e\bar{\gamma}$) with $\delta=2\chi_{\gamma}/(3\bar{\chi}_e(\bar{\chi}_e-\chi_{\gamma}))$, $\tau_c=\hbar/mc^2$ and $\alpha= e^2/\hbar c$. We start from an assumption (which is verified by the result) that in the limit $\bar{\chi}_e\ll 1$, $W_p(\chi_{\gamma})\ll s$, hence the zeros of $s$ corresponding to a growing exponential ($\Gamma=s^{+}$) are given by 
\begin{equation}
\label{eq:lowa0}
\Gamma \simeq \left(2\int_0^{\bar{\chi_e}} d\chi_{\gamma} \frac{d^2P}{dtd\chi_{\gamma}}W_p\right)^{1/2}.
\end{equation}
This integrand in Eq.(\ref{eq:lowa0}) is comprised of an exponential function multiplied by another function. More specifically, the argument of the exponential possess a unique maximum, $\chi_{\gamma,0} = 2\bar{\chi}_e/3$ at which the second derivative of the argument is negative. One can thus evaluate the integral using the Laplace's method: $\int h(x)e^{f(x)}dx \simeq \sqrt{2\pi/|f''(x_0)|}h(x_0)e^{f(x_0)}$ where $x_0$ is the unique maximum. In this case, $f(\chi_{\gamma})= -\delta-8/3\chi_{\gamma}$, $f(\chi_{\gamma,0}) = -16/3\bar{\chi}_e$, $f''(\chi_{\gamma,0})=-54/\bar{\chi}_e^3$, and $h(x_0) = 1/\sqrt{\delta(\chi_{\gamma,0})}=\sqrt{3\bar{\chi}_e}/2$. We obtain for the growth rate in the weak field limit:
\begin{equation}
\Gamma\simeq \frac{1}{5}\sqrt{\frac{{\pi}}{6^{1/2}}}\frac{\alpha}{\tau_c}\frac{\bar{\chi}_e e^{-8/3\bar{\chi}_e}}{\bar{\gamma}}.
\end{equation}
The last step consists in finding how $\bar{\gamma}$ and $\bar{\chi}_e$ depend on $a_0$. In a rotating field mocking the beating of two $1 \mu m$ lasers \cite{Bell_Kirk_2lasers, Elkina_rot, model1bell}, $\vec{a}=a_r[\cos(\omega_0t),\sin(\omega_0t)]$ ($a_r= 2a_0$), it is clear from Eq.(\ref{eq:lowa0}) that $\Gamma \ll \omega_0$ for $\bar{\chi}_e\ll 1$. Thus $\bar{\gamma}$ and $\bar{\chi_e}$ can be approximated by their average values over a laser cycle. The expressions of $\gamma(t)$ and $\chi_e(t)$ can be found in \cite{Elkina_rot} and for $a_r \gg 1$ (neglecting the quantum recoil) one finds $\bar{\gamma}\simeq \langle \gamma \rangle = 4a_r/\pi$ and $\bar{\chi}_e\simeq \langle \chi_e \rangle = a_r^2/a_S$. Although not shown in this article, these estimates are close to the average energy and average quantum parameter of the particles in the simulations. 

{\it Strong field limit:} when $\bar{\chi_e}\gg 1$, $W_{\gamma}$ and $W_p$ have a similar asymptotic expressions and Fedotov \cite {Fedotov_cascade} obtained with intuitive considerations an estimate for the growth rate $\Gamma \sim W_p \sim W_{\gamma}(\bar{\chi}_e,\bar{\gamma})$. Whilst it is not an exact solution, one can verify that this result is somehow consistent with Eq.(\ref{eq:charac}). In this limit, where the recoil cannot be omitted, $\Gamma \gg \omega_0$ and the values of $\bar{\gamma}$ and $\bar{\chi}_e$ can be evaluated as \cite{Fedotov_cascade} $\bar{\gamma}\sim \gamma(t=W_{\gamma}^{-1})\simeq\mu^{3/4}\sqrt{a_S}$ and $\bar{\chi}_e\sim \chi_e(t=W_{\gamma}^{-1})\simeq 1.24\mu^{3/2}$ with $\mu= a_r/(\alpha a_S)$. Taking the characteristic energy for photon emission at the moment $W_{\gamma}^{-1}$ has been proven to be a valuable and accurate prediction \cite{Elkina_rot}. Unfortunately it is not possible to obtain a simple analytical expression for the growth rate in such limit and we resorted to a numerical computation of Eq.(\ref{eq:charac}).

These two asymptotic limits are the generalization of the growth rate obtained in the ideal model for $r\sim 1$ and $r \gg 1$. The case $r \ll 1$ is not physically relevant since photon emission is always more probable than pair emission.

\section{Physical setup and laser parameters}
\label{section_para}

\begin {table}[!t]
%\centering
\caption{Growth rate for plane wave and for laser pulses. The growth rate is measured in units of the plane wave/ laser pulse frequency $\omega_0$. PW stands for plane wave whereas pulse stands for simulation performed with realistic laser pulses whose parameters are described in the text.}
\label{plane_pulse}\centering
\begin {tabular}{c|c|c|c|c}
\hline
$\bold{a_0}$ & \bf 1000 &  \bf 1500 & \bf 3000 & \bf 5000 \\
\hline
type & pulse / PW & pulse / PW & pulse / PW & pulse / PW \\ 
$\Gamma_{cw-cp}$ & 0.85 / 0.82 & 1.35 / 1.39 & 2.4 / 2.39 & 3.35 / 3.38 \\ 
$\Gamma_{lp-lp}$ & 0.28 / 0.25 & 0.56 / 0.52 & 0.85 / 0.82 & 1.2 / 1.05 \\ 
$\Gamma_{cw-cw}$ & 0.13 / 0.14 & 0.17 / 0.15 & 0.2 / 0.19 & 0.25 / 0.23 \\
\end {tabular}
\end {table}

The laser parameters we chose are inspired on the typical parameters expected in future laser facilities such as Vulcan or ELI \cite{ELI,ELI_WhiteBook,Vulcan}, 10 PW peak-power system, 100J-1kJ, 30-60 fs and a focal spot that could be as small as the micron size. We have also pushed the parameters in order to make the bridge between different regimes: the onset of QED characterized by $\chi \ll 1$ \cite{model1bell, Bell_Kirk_2lasers, Bell_Kirk_MC} and the full QED dominated regime for $\chi \gg 1$, that have been explored in prior studies \cite{Fedotov_cascade, Nerush_laserlimit, Elkina_rot, Bashmakov}. 

One of the objectives of these future facilities is indeed to focus these ultra intense lasers to a micron spot size and this can be probably achieved by using adequate optics. Nonetheless, from a theoretical point of view, focusing a laser pulse to a given waist requires knowing the self-consistent shape of the pulse far away from the focus point. The seminal article on electromagnetic beams from L.W. Davis \cite{Davis} shows how to construct light paraxial beams whose formal solution employs an expansion in power of $W_0/z_r$, where $W_0$ is the beam waist and $z_r$ the diffraction length. The well-known solution used in the literature for Gaussian beams requires $W_0 \ll z_r$ or equivalently $\lambda_0 \ll W_0$ ($\lambda_0$ being the central wavelength of the laser beam). Hence, a paraxial beam is not an accurate solution for a beam which is aimed to be focused at $\lambda_0 \simeq W_0$. Furthermore, it has also been proven \cite{Salamin} that focusing a laser beam to the diffraction limit requires inclusion of terms of fifth order in the diffraction parameter $W_0/z_r$ in the description of the associated fields. Another conclusion drawn from this latter article is that the electron dynamics in tightly focused beam is not adequately described by the plane-wave approximation because of the extra components of the field that must be considered to satisfy $\vec{\nabla}\cdot\vec{E}=0$ and $\vec{\nabla}\cdot\vec{B}=0$ in vacuum. 

In our simulations, all laser pulses have a $\lambda_0 = 1 \mu \mathrm{m}$ central wavelength, and the same spatio-temporal envelope functions, with differences in the fast-oscillating components that will be presented separately for different polarizations. The envelope function is transversally a Gaussian  with a focal spot of 3.2 $\mu \mathrm{m}$, while the temporal profile is given by $10\tau^3-15\tau^4+6\tau^5$, $\tau=t/\tau_0$ for $t\leq \tau_0$ and $\tau=2\tau_0-t$ for $\tau_0< t\leq 2 \tau_0$, where $\tau_0=32~$fs is the pulse duration at FWHM in the fields. The focal spot of 3.2 $\lambda$ represents a compromise where we can ensure that the laser intensity at the focus is the one wanted and that in the region of the focus the structure of the fields is close to a plane wave. 
 
The laser pulses are initialized 20 $\mu \mathrm{m}$ away from one another. The focal plane for both lasers is located at half distance between their envelope centers. 100 test electrons are placed in the focal plane to seed the cascade in 3D (10 electrons in each transverse direction, they occupy the area of 1 $c^2/\omega_0^2$). 2D simulations were seeded with one 100 electrons in the transverse direction over a length of $c/\omega_0$. We have also tested the seeding with one electron for 2D and 3D simulations where the electron was located at the very center of the focal plane. The simulation box is composed of $3000\times 1200$ cells and $3000\times 1200\times 1200$ cells for 2D and 3D, respectively. The spatial resolution is $dx=dy=dz=0.1 ~c/\omega_0$ and after extensive convergence tests we have chosen $dt= 0.001~ \omega_0^{-1}$ ($\omega_0=k_0c=2\pi c/\lambda_0$). 

In order to put the model exposed in the previous section to the test, we have also conducted a series of simulations with a uniform rotating electric field. The rotating field is imposed as a external field and is turned on during the entire time of the simulation and has the following structure: $E_x = a_0\cos(\omega_0t)$ and $E_y = a_0\sin(\omega_0t)$. The spatial uniformity of the electric field allows to use a small simulation box with periodic boundary conditions. Nonetheless, the time resolution is still conditioned by the pair and photon characteristic emission time and we have thus kept $dt= 0.001~ \omega_0^{-1}$. 

\section{Discussion and Conclusion}

\begin{figure}
\includegraphics[width=0.5\textwidth]{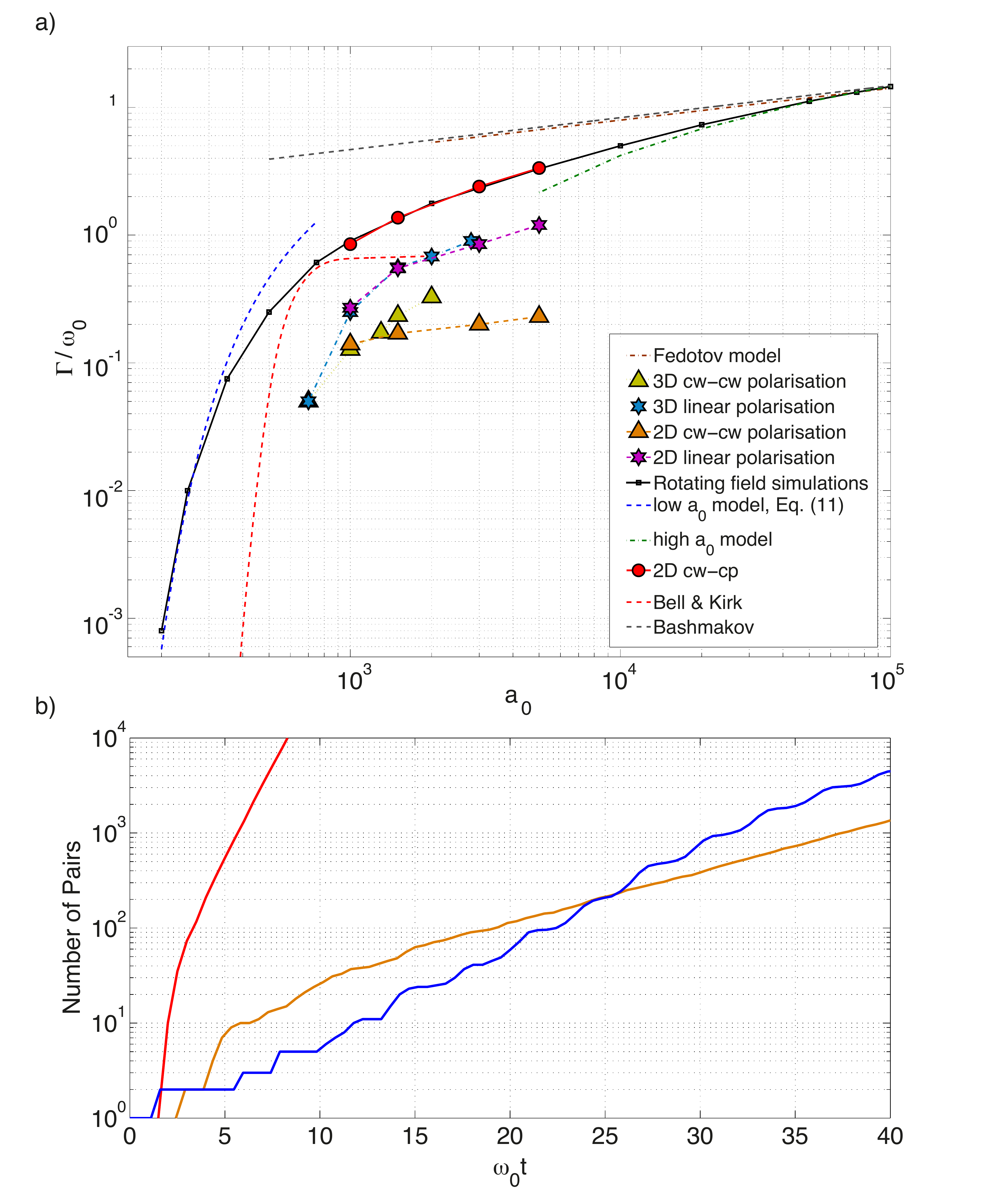}
\caption{a) Growth rate as a function of $a_0$ for different laser polarization. The low $a_0$ model corresponds to the Eq. \eqref{eq:lowa0} and the high $a_0$ model corresponds to the numerically integrated Eq. \eqref{eq:charac} with $\bar{\gamma}=\mu^{3/4}\sqrt{a_S}$ and $\bar{\chi}_e= 1.24\mu^{3/2}$. b) onset of QED cascades for the three setups for $a_0=1000$.} 
\label{growth_rate}
\end{figure}

As shown recently by Jirka {\it et al.}\cite{Jirka}, the spatial structure of the cascading plasma is essential to understand where the pairs are produced. Additionally, the growth rate of the cascade is the important macroscopic quantity that tells how the number of pairs rises in the interaction region. As a matter of fact, as we will see in this section, the growth rate of the cascade depends on the polarization of the lasers which produce different field structures. Additionally, the growth rate of the cascade can only be determined in a unambiguous manner when the density of the pair plasma is relativistically under dense such that the self-generated field remains negligible compared to the Òexternal fieldsÓ provided by the overlap of the two lasers.  The case of relativistically over dense pair plasmas is of high relevance for laser absorption and this problem has been addressed with QED-PIC showing that significant absorption can be achieved for $I >  10^{24}$ W/cm$^2$ and pulses of few 10's of fs as first demonstrated by Nerush \cite{Nerush_laserlimit} and later by Grismayer \cite{GrismayerPOP}. As mentioned previously, the focal spot of both lasers is 3.2 microns, which is still in the paraxial approximation for a one-micron wavelength and thus ensures that at the focus, the field structure of the laser can be described by a plane wave. We have actually performed several simulations with pure plane standing waves and compared the growth rate of the cascade with situations where realistic pulses were used without significant differences, as can be seen in Table(\ref{plane_pulse}). It should be stressed that we measure the growth rate in realistic setup when the two lasers are around the point of full overlap in order to make sensible comparisons with ideal setups (such as the uniform rotating field).
 
Figure \ref{growth_rate} shows the growth rate for different configurations as a function of $a_0$. The simulation results for the pure rotating electric field configuration are displayed in black squares/lines. The growth rate given by Eq.(\ref{eq:lowa0}) and the numerical solution of Eq.(\ref{eq:charac}) depicted respectively in the blue and green dashed lines are in good agreement with the rotating field simulation results in the limit of their validity ($a_0 \ll 10^3$ for $\bar{\chi}_e \ll 1$ and $a_0 \gg 10^3$ for $\bar{\chi}_e \gg 1$). As expected, the growth rates in the cw-cc setup match those of the rotating field configuration. This growth rate is the highest of all the three configurations for a fixed $a_0$. In order to underline the improvement of our model compared to previous ones, we have also plotted the growth rate derived by Bell \& Kirk \cite{model1bell} in dashed red, the rate derived by Bashmakov \cite{Bashmakov} in brown dashed and the result given by Fedotov \cite{Fedotov_cascade}. In the weak field regime, the rate coming from the model of Bell \& Kirk is low in comparison with our analytical and numerical study. Even though their model and ours are not in essence different, one of the reason of the mismatch is that the successive approximations made by Bell \& Kirk led to a underestimate of the photon optical depth. In the strong field regime, both prior models \cite{Bashmakov, Fedotov_cascade}, which possess the same scaling, appear to be valuable predictions for extreme intensities ($a_0 \gg 10^4$). The small discrepancy, that occurs at lower intensity, with our analytical results (curve in green dashed) results from the approximation $\Gamma \sim W_p \sim W_{\gamma}$. As discussed before, whilst this ordering is appropriate, numerical factors (that are neglected) are sufficient to cause a departure from the numerical results.  

The lp-lp setup has a growth rate lower than the cw-cc configuration, but higher than the cw-cw. Figure \ref{growth_rate}b) also confirms the $2\omega_0$ oscillating component of the lp-lp growth rate. There is no appreciable difference in the growth rate of the 2D and 3D simulations for linearly polarized lasers. The lowest growth rate is attributed to the cw-cw configuration, and this is also in agreement with our previous analysis. Finite size Gaussian laser pulses can provide an initial non zero $p_x$ in order to seed the cascade. The gradients of the intensity that provide the ponderomotive force are more pronounced in 3D (they affect a higher percentage of the particles), and therefore the growth rate in this configuration becomes slightly higher in 3D than in 2D simulations. This configuration is robust because there are no special favorable locations for the cascade seeding. On the contrary, the seeding with electrons of the setup cw-cc turns out to be difficult. The reason is that an efficient growth happens only in the regions around the maximum of the electric field. By starting a cascade with only a few electrons, it is not guaranteed that they will enter such a region. This is precisely why there is no 3D data for the cw-cc configuration in Fig. \ref{growth_rate}a): the cascade has not started below a0 = 2000 even though the same initial conditions were used as in Setup 1 and Setup 2. Similar conclusions have been drawn by Jirka \cite{Jirka} who recommends the linear polarization setup in order to maximize the number of pairs created.  

\begin {table}[!t]
%\centering
\caption{Number of pairs per initial electron obtained for linearly polarised laser pulses with $W_0 = 3.2\mu m$}
\label{cascade_yield}\centering
\begin {tabular}{ccccc}
\hline
$\bold{a_0}$ & \bf dimension &  \bf seeding & \bf pairs/$\bold{e^{-}}$ & \bf cascade \\
\hline
400 & 2D & 100 electrons & 0.01 & \ding{55} \\ 
500 & 3D & 100 electrons & 0.03 & \ding{55} \\ 
700 & 2D & 100 electrons & 10   & \checkmark \\
700 & 3D & one electron  & 30 & \checkmark \\
\end {tabular}
\end {table}

A crucial question addressed by Fedotov {\it et al.}\cite {Fedotov_cascade} concerns the cascade threshold, that according to the latter author, is around $I_{th} >  2.5 \times 10^{25}$ W/cm$^2$ or $a_r > \alpha a_S$. Fundamentally speaking, there is no threshold for cascading process since the growth rate never vanishes (even tough being exponentially reduced for lower $a_0$'s). Notwithstanding, a very low growth rate requires a very long laser duration as well as a gigantic spot size in order to show evidence of the cascade. A convincing definition of the threshold can therefore be the minimum growth rate required to observe a significant amount of pairs produced, i.e., few e-folding of the cascade, $\Gamma \tau_c \gtrsim 1$, where $\tau_c$ is the characteristic time during which the process occurs. Fedotov {\it et al.}\cite {Fedotov_cascade}, who established a model valid in the strong field limit, identified two characteristic times: the acceleration time which is the time required to reach $\chi \simeq 1$ and the escaping time which is the time of stay of a particle (pairs or photon) in the laser pulse. The minimum growth rate lies probably in the the weak field limit where $\chi < 1$, so the acceleration time is not necessary. However, the escaping time is still relevant in our case and can be generalized as $t_{esc} \sim W_0 / c$. Using Eq.\ref{eq:lowa0}, we obtain a new threshold given by
\begin{equation}
\label{eq:threshold}
a_0 > \sqrt{\frac{2 a_S}{3\log(\frac{\pi a_0\alpha W_0c}{8\omega_0})}}.  
\end{equation}
For $W_0 = 3\mu m$, $a_0 > 310$ or $I >  2.7\times 10^{23}$ W/cm$^2$, which is two orders of magnitude lower than the threshold derived initially by Fedotov {\it et al.}\cite {Fedotov_cascade}. One should justly notice that in a recent publication \cite{Fedotov_threshold}, the same author recognizes that the initial criteria, $a_r > \alpha a_S$, in fact overestimates the actual cascade threshold for three main reasons: i) simulations show cascades development a lower intensity, ii) the escaping time that may be larger and iii) that pairs can be only created when $\chi_{\gamma} \gtrsim 1$. We shall now consider what is the threshold for cascading process taking into account the laser parameters of future facilities. The laser power should theoretically rise up to 10 PW. For a focal spot close to the diffraction limit, $W_0 \simeq \lambda_0$, the intensity could then reach $I \simeq  10^{24}$ W/cm$^2$. Applying the same criteria for the threshold, $\Gamma > 1/t_{esc}$ and using the results of Fig.\ref{growth_rate}a), we find that in case of the setup cw-cc, $a_0 > 400$ and for the setup lp-lp, $a_0 > 800$, which corresponds in both cases to intensities lower than $I =  10^{24}$ W/cm$^2$. We would like to warn the reader that these numbers are conjectural since we have not performed simulations for focal spots close to the diffraction limit. Nonetheless these numbers appear to be on the same order as the threshold given by Gelfer\cite{Gelfer}. If the focal spot was taken to be larger such as the one we chose in this work, $W_0 = 3.2\mu m$, for linearly polarized lasers (which are the more likely to be delivered), observing a cascade will require $a_0 > 700$ or $I_{th} >  6.8\times 10^{23}$ W/cm$^2$. This threshold can be verified in Table(\ref{cascade_yield}) for the laser pulses parameters considered in this study. The criteria for the threshold is $\Gamma W_0/c >1$, which implies that every single electron participating to the cascade will produce at least produce three new electrons during the interaction of the two laser pulses. This intensity corresponds to a peak power of $P=\pi W_0^2*I_{th}/2\simeq 100$ PW. 

In summary, the efficiency of QED cascades has been studied for three different laser intensities and configurations in 2D/3D simulations. Whereas the setup 3 seems to be promising due to an unquestionably higher growth rate for a fixed $a_0$, the seeding of this latter configuration proves to be problematic. The setup 1 and 2 are more recommendable to ensure the take-off the cascade. Using the growth rates of Fig.\ref{growth_rate}a), we predict that the cascading process should start around $I >  7\times10^{23}$ W/cm$^2$ for a focal spot above the diffraction limit and 30 fs lasers. With an electron seeding composed of a micron-size cryogenic hydrogen target, a relativistic critical density pair plasma $n_{rc}$ can be created for the parameters expected for ELI \cite{ELI_WhiteBook} ($I >  10^{24}$ W/cm$^2$ for pulses of few 10's of fs). Once the plasma reaches the density $n_{rc}$, the laser starts to be efficiently converted into gamma rays and one approaches the condition to create a laboratory gamma-ray pulsar \cite{Gruzinov,Nerush_laserlimit,GrismayerPOP}. This will be explored in future publications.

\begin{acknowledgments}
This work is supported by the European Research Council (ERC-2015-AdG Grant 695088) and FCT (Portugal) Grants  SFRH/BD/62137/2009 and SFRH/IF/01780/2013. We acknowledge PRACE for awarding access to resource SuperMUC based in Germany at Leibniz research center. Simulations were performed at the Accelerates cluster (Lisbon, Portugal), and SuperMUC (Germany).
\end{acknowledgments}

%\bibliography{cascade.bib}
%\bibliography{cascade_model.bbl}

%\end{thebibliography}%


%

\end{document}